\pdfoutput=1
\documentclass[journal]{IEEEtran}
\usepackage{amsmath,amssymb}
\usepackage{graphicx}
\usepackage{cite}
\usepackage{booktabs}
\usepackage{url}
\usepackage{xcolor}
\newcommand{\bH}{\mathbf{H}}
\newcommand{\hH}{\hat{\mathbf{H}}}
\newcommand{\bQ}{\mathbf{Q}}
\newcommand{\bV}{\tilde{\mathbf{V}}}
\newcommand{\hh}{\hat{\mathbf{h}}}
\DeclareMathOperator*{\median}{median}
\begin{document}

\title{Detecting Transmit Precoding in Commodity Wi-Fi Channel State Information and Its Effect on Measured Channel Statistics}

\author{Raghuram~C.~Kamath, Carlos~Aldana, and Kavya~Guruswamy%
\thanks{This work involved human subjects in its research. The authors confirm that all human subject research procedures and protocols are exempt from review board approval. All participants gave informed consent.}%
\thanks{R.~C.~Kamath, C.~Aldana, and K.~Guruswamy are with Meta Platforms, Inc., Sunnyvale, CA, USA (e-mail: raghuck@meta.com; caldana@meta.com; kavyag@meta.com).}}

\markboth{IEEE Wireless Communications Letters}{Kamath \MakeLowercase{\textit{et al.}}: Detecting Transmit Precoding in Commodity Wi-Fi CSI}

\maketitle

\begin{abstract}
Commodity Wi-Fi channel state information (CSI) is increasingly used to measure channels, but for beamformed multi-stream frames it describes the precoded channel. A blind test based on the stream columns' near-orthogonality and power ordering detects eigen-beamforming without calibration or packet alignment. Simulated false-alarm rates are at most $2.9\times10^{-3}$; precoding scales per-entry RMS delay spread by 0.3--5.7 but nearly preserves total power. At two commercial access points, 880 of 896 two-stream frames tested as precoded, including 601 of 605 without a beamforming flag; at a third with beamforming off, none of 3223 did.
\end{abstract}


\section{Introduction}
\IEEEPARstart{C}{ommodity} Wi-Fi receivers export per-subcarrier channel state information (CSI) \cite{halperin2011tool,gringoli2019free,gringoli2021axcsi,jiang2022picoscenes}. The CSI drives a large body of sensing work \cite{ma2019wifi,du2025overview,ropitault2024sensing,ieee80211bf} and is increasingly used as a low-cost sounder: power delay profiles from stitched bands \cite{xie2015precise}, delay-synchronous wideband sounding \cite{yamamoto2025delay}, and over-the-air referencing \cite{yamamoto2026radar}. Unlike a sounder's output, the CSI carries well-characterized receiver artifacts \cite{jiang2022picoscenes,portner2026same,cominelli2023exposing,cominelli2025glimpse,tonini2025towards}.

The transmitter adds a second layer that has received less attention. For high-throughput (HT) and later physical-layer protocol data units (PPDUs) with more than one spatial stream, the transmitter maps streams to antennas through a spatial mapping matrix $\bQ(k)$. This matrix may be an eigen-beamformer computed from the receiver's compressed feedback \cite{ieee80211_2020,ieee80211ax,ieee80211be,perahia2013next}, and the CSI then estimates $\bH(k)\bQ(k)$. Beamforming feedback has been used as a sensing signal \cite{meneghello2023sharp,kondo2022bidirectional,haque2025beamsense,haque2023wibfi,li2024efficient}, transmitter-side CSI obfuscation has been proposed for privacy \cite{cominelli2021csi,cominelli2022antisense,cominelli2024physical}, and manipulated feedback can corrupt multi-user precoding \cite{meneghello2025break}. Cellular receivers can blindly detect the precoding matrix index of an interfering signal within a known codebook \cite{kwon2016blind}; the test here needs no codebook and serves measurement validity. We are not aware of work that quantifies how routine transmit beamforming alters CSI-derived channel statistics, or that gives a measurement campaign a way to tell whether its CSI was precoded.

\begin{figure}[!t]
\centering
\includegraphics[width=\columnwidth]{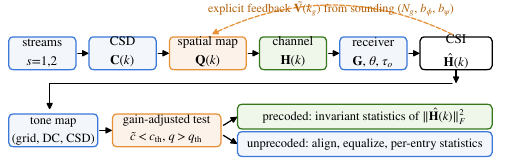}
\caption{Signal model (top) and processing chain (bottom). The test minimizes the column coherence over the receive-chain gain ratio.}
\label{fig:model}
\end{figure}

\section{Signal Model}
For one PPDU with $N_s$ streams and $N_r$ receive chains, the CSI at subcarrier $k$ is (Fig.~\ref{fig:model})
\begin{equation}
\hH(k)=\beta_k\,\mathbf{G}\,\bH(k)\,\bQ(k)\,\mathbf{C}(k)+\mathbf{N}(k),
\label{eq:model}
\end{equation}
where $\beta_k=\alpha e^{j(\theta-2\pi k\Delta f\tau_o)}$, $\bH(k)\in\mathbb{C}^{N_r\times N_t}$ is the propagation channel including antennas, $\bQ(k)\in\mathbb{C}^{N_t\times N_s}$ is the spatial mapping, $\mathbf{C}(k)=\mathrm{diag}(e^{-j2\pi k\Delta f\delta_s})$ holds the per-stream cyclic shifts, $\mathbf{G}=\mathrm{diag}(g_1,\dots,g_{N_r})$ holds the complex receive-chain gains, $\alpha$, $\theta$, and $\tau_o$ are the common amplitude, phase, and timing offset of the PPDU, $\Delta f$ is the subcarrier spacing, and $\mathbf{N}(k)$ is estimation noise. With direct mapping, $N_t=N_s$ and $\bQ=\mathbf{I}$. With spatial expansion, $\bQ$ is fixed up to per-antenna cyclic shifts. With beamforming, $\bQ(k)=\bV(k_g)$, where $k_g$ is the feedback subcarrier of the group of $N_g$ subcarriers that contains $k$. $\bV$ is reconstructed from Givens angles, quantized with $(b_\phi,b_\psi)$ bits, of the right singular vectors of the beamformee's estimate $\mathbf{G}_s\bH(k_g)$ at the sounding instant \cite{ieee80211_2020,ieee80211ax,perahia2013next}. The compressed form makes the last row of $\bV$ real and non-negative. $\bQ(k)$ therefore carries a column phase that follows that row across subcarriers and is piecewise constant across groups.

With fresh, unquantized feedback and $\mathbf{G}=\mathbf{G}_s$, the thin singular value decomposition $\mathbf{G}\bH=\mathbf{U}\boldsymbol{\Sigma}\mathbf{V}^H$ gives
\begin{equation}
\hH\propto\mathbf{U}\boldsymbol{\Sigma}\,\mathbf{C},\qquad \hh_s\perp\hh_{s'}\ (s\neq s'),\qquad \|\hh_1\|\ge\|\hh_2\|,
\label{eq:svd}
\end{equation}
where $\hh_s$ is column $s$ of $\hH$. An access point (AP) that beamforms implicitly, from its own estimate of the uplink channel, produces the same structure without feedback quantization. Quantization, grouping, and aging over the feedback delay perturb \eqref{eq:svd}. We model aging as $\bH_d=\rho\bH_s+\sqrt{1-\rho^2}\,\mathbf{W}$, with $\mathbf{W}$ independent of and distributed as $\bH_s$, or as its diffuse part under line of sight (LOS), and $\rho=J_0(2\pi f_D\Delta t)$.

\emph{Remark (non-HT frames):} Beacons and other non-HT PPDUs are not precoded, but may be sent from several chains with cyclic shifts. For uncorrelated equal-power chains, the average power delay profile (PDP) is a mean of shifted copies, and
\begin{equation}
\tau_{\mathrm{rms,meas}}^2=\tau_{\mathrm{rms}}^2+\sigma_\delta^2,
\label{eq:cdd}
\end{equation}
where $\sigma_\delta$ is the RMS of the shift set: 100, 81.6, and 55.9~ns for the two- to four-chain pre-HT sets \cite{ieee80211_2020}.

\section{Blind Precoding Test}
\subsection{Statistics and null distribution}
Over the $K$ valid subcarriers of a PPDU with $N_s\ge2$, define, for a receive scaling $\mathbf{D}=\mathrm{diag}(d,1)$ ($N_r=2$; further chains need one ratio each and a lower $c_{\mathrm{th}}$), the column coherence and the power-ordering fraction
\begin{gather}
c(k;\mathbf{D})=\frac{|\hh_1^H\mathbf{D}^2\hh_2|}{\|\mathbf{D}\hh_1\|\,\|\mathbf{D}\hh_2\|},\qquad \tilde c=\min_{\mathbf{D}}\median_k c(k;\mathbf{D}),\label{eq:stats}\\
q=\frac{1}{K}\sum_k \mathbf{1}\!\left[\|\hat{\mathbf{D}}\hh_1(k)\|>\|\hat{\mathbf{D}}\hh_2(k)\|\right],\label{eq:q}
\end{gather}
where $\hat{\mathbf{D}}$ attains the minimum over $20\log_{10}d\in[-15,15]$~dB in 0.5-dB steps. Both statistics are invariant to $\alpha$, $\theta$, $\tau_o$, the receive-chain phases $\arg g_r$, and $\mathbf{C}(k)$, which act as row or column phases; $c$ is also invariant to per-stream power scaling, and the minimization absorbs the receive-chain gain ratio within the searched range. The test therefore needs no packet alignment, cyclic-shift removal, or gain or phase calibration. This matters because $\bV$ comes from $\mathbf{G}_s\bH$ at sounding, whereas a data frame may arrive with gains $\mathbf{G}$: for orthonormal left singular vectors $\mathbf{u}_1,\mathbf{u}_2$ of $\mathbf{G}_s\bH$ and $\mathrm{diag}(d_1,d_2)=\mathbf{G}\mathbf{G}_s^{-1}$, their inner product becomes $(|d_1|^2-|d_2|^2)\,u_{11}^*u_{12}$, so $c$ can rise to $|x-1|/(x+1)$, $x=|d_1/d_2|^2$; per-chain rescaling of the CSI, such as to the received signal strength indicator (RSSI), acts alike. The test requires $N_s\ge2$: a single-stream beamformed PPDU cannot be identified from its CSI. Multi-user and deliberately randomized precoding \cite{cominelli2024physical} need not produce orthogonal columns and are outside its scope.

For an unprecoded PPDU in independent and identically distributed (i.i.d.) Rayleigh fading and $\mathbf{D}=\mathbf{I}$, $\hh_1$ and $\hh_2$ are independent isotropic vectors in $\mathbb{C}^{N_r}$, so $c(k)^2\sim\mathrm{Beta}(1,N_r-1)$ \cite{love2003grassmannian,mukkavilli2003beamforming}. In general $P[c(k)\le x]=1-(1-x^2)^{N_r-1}$; for $N_r=2$ this is $x^2$, with median $1/\sqrt2$. The norm and the direction of an isotropic Gaussian vector are independent, so $c$ and $q$ are independent under the null, and $q$ has median $1/2$. Correlation, an LOS component, and noise break this independence and are evaluated by simulation. With $M\approx B/B_{c,50}$ approximately independent subbands, where $B$ is the bandwidth and $B_{c,50}$ is the 50\% coherence bandwidth, the decision $\{\tilde c<c_{\mathrm{th}}\}\cap\{q>q_{\mathrm{th}}\}$ has
\begin{equation}
P_{\mathrm{FA}}\approx P\!\left[\mathcal{B}(M,c_{\mathrm{th}}^2)\ge\lceil M/2\rceil\right]P\!\left[\mathcal{B}(M,\tfrac12)>q_{\mathrm{th}}M\right],
\label{eq:pfa}
\end{equation}
where $\mathcal{B}(M,p)$ is a binomial variable. For $c_{\mathrm{th}}=0.45$ and $q_{\mathrm{th}}=0.9$, \eqref{eq:pfa} gives $1.2\times10^{-2}$, $2.3\times10^{-4}$, and $1.1\times10^{-5}$ for $\tau_{\mathrm{rms}}=15$, 30, and 50~ns ($M\approx4$, 8, and 14 from the simulated $B_{c,50}$ of 19, 9.2, and 5.5~MHz over $B=78$~MHz). Simulation at $\mathbf{D}=\mathbf{I}$ ($4\times10^4$ PPDUs per case, i.i.d., $N_t=2$, 30~dB) gives $1.3\times10^{-3}$ at 15~ns and none at 30 or 50~ns (95\% bound $7.5\times10^{-5}$), so \eqref{eq:pfa} is a rough, conservative guide. The minimization over $\mathbf{D}$ doubles the 15-ns rate to $2.6\times10^{-3}$ and gives $10^{-4}$ at 30~ns and none at 50~ns; the rates quoted below are simulated with it. Correlation and a LOS component raise $c$ under the null and lower $P_{\mathrm{FA}}$; unequal receive gains no longer do. Under precoding, \eqref{eq:svd} drives $c\to0$ and $q\to1$.

\subsection{Monte Carlo evaluation}
\begin{figure}[!t]
\centering
\includegraphics[width=\columnwidth]{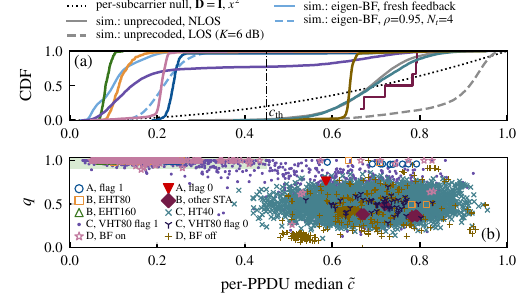}
\caption{(a) Cumulative distribution function (CDF) of the gain-adjusted median column coherence $\tilde c$: per-subcarrier null at $\mathbf{D}=\mathbf{I}$, simulated unprecoded and beamformed channels (gray, light blue), and measured PPDUs (colors as in (b); D: both sessions). (b) Measured $\tilde c$ versus $q$ (shaded: decision region; BF: beamforming; STA: station).}
\label{fig:test}
\end{figure}

We simulate an 80~MHz channel on 501 subcarriers at $\Delta f=156.25$~kHz, the grid of the measured CSI in Section~V. Each PDP is exponential with 5~ns taps and $\tau_{\mathrm{rms}}\in\{15,30,50,100\}$~ns, the RMS delay spreads of TGn models B--E \cite{erceg2004tgn}, which TGax retains for indoor channels \cite{liu2014tgax}. Fading is either non-LOS (NLOS) i.i.d.\ Rayleigh or LOS with $K=6$~dB, half-wavelength linear arrays, and random departure and arrival angles. The array sizes are $N_r=N_s=2$ and $N_t\in\{2,4\}$. Unprecoded frames use direct mapping ($N_t=2$) or spatial expansion ($N_t=4$). Beamformed frames use single-user feedback with $N_g\in\{4,16\}$ (every 2 or 8 grid subcarriers), $(b_\phi,b_\psi)\in\{(6,4),(4,2)\}$ \cite{ieee80211ax}, and $\rho\in\{1,0.99,0.95,0.9\}$. The signal-to-noise ratio (SNR), the mean entry power over the noise variance, is 30~dB, with 1500 PPDUs per case. A second set varies the SNR (10, 20, 30~dB) and adds exponential correlation of 0.5 or 0.9 between adjacent antennas at both ends (1000 PPDUs per case). The simulated null medians of $\tilde c$ are 0.67--0.70 for NLOS, below $1/\sqrt2$ because of the minimization, and 0.89--0.92 for LOS (Fig.~\ref{fig:test}(a)). Table~\ref{tab:det} lists detection rates.

\begin{table}[!t]
\centering
\caption{Simulated Fraction of PPDUs Declared Precoded ($c_{\mathrm{th}}=0.45$, $q_{\mathrm{th}}=0.9$, 30~dB, $\tau_{\mathrm{rms}}=15$--100~ns, NLOS and LOS; False Alarm: 15~ns, i.i.d., $4\times10^4$ PPDUs)}
\label{tab:det}
\footnotesize
\begin{tabular}{@{}lcc@{}}
\toprule
Transmission & $N_t=2$ & $N_t=4$\\
\midrule
Unprecoded (false alarm) & $2.6\times10^{-3}$ & $2.9\times10^{-3}$\\
Beamformed, $\rho=1$, $N_g=4$, $(6,4)$ & $\ge0.999$ & $1.000$\\
Beamformed, $\rho=0.95$, $N_g=4$, $(6,4)$ & 0.67--0.97 & 0.97--1.00\\
Beamformed, $\rho=0.95$, $N_g=16$, $(4,2)$ & 0.00--0.51 & 0.80--0.99\\
Beamformed, $\rho=0.90$, $N_g=16$, $(4,2)$ & 0.00--0.21 & 0.47--0.81\\
\bottomrule
\end{tabular}
\end{table}

The test detects precoding with fresh feedback in at least 99.9\% of PPDUs and stays reliable for $N_t=4$ under moderate aging, but with $N_t=2$ and stale, coarse feedback it misses most precoded frames. The aging term $\sqrt{1-\rho^2}\,\mathbf{W}\bQ$ has the same power for any $N_t$, but the weaker eigenmode of a $2\times2$ channel is often in a deep fade, where that term dominates and drives $c$ toward its null value. If $\bV$ comes from $\mathbf{G}_s\bH$ and one chain's gain then changes by 6 or 10~dB, fresh feedback still gives $P_D\ge0.99$, against at most 0.26 and 0.01 at $\mathbf{D}=\mathbf{I}$. In the second set, false alarms appeared only at 15~ns; rerun with $4\times10^4$ PPDUs, these cases give $1.6$--$2.9\times10^{-3}$ (i.i.d., 20--30~dB), $1.3$--$2\times10^{-4}$ (10~dB), and at most $7.3\times10^{-4}$ with LOS or correlation 0.5, so the 15-ns i.i.d.\ case bounds the rest, whose zero counts in 1500 or 1000 PPDUs give only $2\times10^{-3}$ and $3\times10^{-3}$. Correlation and LOS raise the null median of $\tilde c$ to 0.79--0.99, lowering false alarms and detection. Fresh feedback gives $P_D\ge0.98$ at 20 and 30~dB for i.i.d., LOS, and correlation-0.5 channels, 0.72--0.89 ($N_t=2$, i.i.d.) and 0.51--0.54 ($N_t=4$, LOS) at 10~dB, and with correlation 0.9 ($N_t=2$) zero at 10~dB and 0.36--0.66 at 30~dB. A miss returns the frame to the unprecoded class, but its statistics remain those of $\bH\bQ$ (Section~IV), so a campaign should decide per capture, for instance on the median of $\tilde c$.

\section{Effect on Channel Statistics}
We compute per-entry statistics as a CSI-based sounder would: the RMS delay spread $\tau_{\mathrm{rms}}$ from the PDP averaged over PPDUs and receive chains (Hann window, threshold 25~dB below the peak, window response removed in quadrature), and $B_{c,50}$ as the lag at which the magnitude of the frequency autocorrelation falls to 0.5 \cite{molisch2011wireless}. A frequency-flat $\bQ$ would keep the PDP shape of i.i.d.\ entries, but eigen-beamforming makes $\bQ(k)$ follow the eigenvectors of $\bH(k)$, carry the column-phase normalization of the compressed feedback, and hold for $N_g$ subcarriers, and each adds frequency structure absent from the channel.

\begin{figure}[!t]
\centering
\includegraphics[width=\columnwidth]{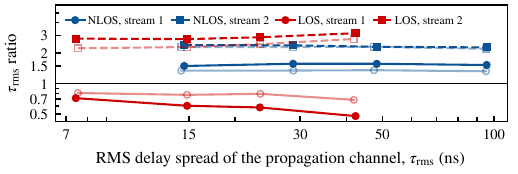}
\caption{Simulated ratio of the per-entry RMS delay spread of the precoded channel to that of the propagation channel, per stream (filled: $N_t=4$; hollow: $N_t=2$; $N_g=4$, $(6,4)$, $\rho=0.99$).}
\label{fig:bias}
\end{figure}

Fig.~\ref{fig:bias} and Table~\ref{tab:bias} show the result with the decoded feedback as is. In NLOS, $\tau_{\mathrm{rms}}$ grows by 1.33--1.65 for stream~1 and 2.2--3.3 for stream~2; in LOS, stream~1 absorbs the dominant path (0.45--0.84) and stream~2 grows by 2.2--5.7. Column phases are free per subcarrier; keeping them continuous gives 0.81--0.95 (stream~1) and 1.1--2.8 (stream~2) in NLOS and 0.29--0.69 and 1.2--5.5 in LOS. Averaged over channels, the streams' delay-spread ratio is 1.5--2.5 in NLOS and up to 11 in LOS, although both columns see the same channel; for one realization it spans 1.1--3.2 in NLOS against 0.75--1.35 unprecoded (10th--90th percentiles). The bias cannot be undone without $\bQ$, which the receiver does not observe.

\begin{table}[!t]
\centering
\caption{Simulated Ratio of Precoded to Propagation-Channel Statistics ($\tau_{\mathrm{rms}}=15$--100~ns; $\rho=0.99$, $(6,4)$, $N_g=4$ and $\rho=0.95$, $(4,2)$, $N_g=16$)}
\label{tab:bias}
\scriptsize
\setlength{\tabcolsep}{3.2pt}
\begin{tabular}{@{}lccccc@{}}
\toprule
 & \multicolumn{2}{c}{$\tau_{\mathrm{rms}}$ ratio} & \multicolumn{2}{c}{$B_{c,50}$ ratio} & $\|\hH\|_F^2$ error\\
\cmidrule(lr){2-3}\cmidrule(lr){4-5}
Case & stream 1 & stream 2 & stream 1 & stream 2 & median (dB)\\
\midrule
NLOS, $N_t=2$ & 1.33--1.42 & 2.21--3.30 & 0.71--0.82 & 0.33--0.39 & 0.00\\
NLOS, $N_t=4$ & 1.49--1.65 & 2.28--3.34 & 0.62--0.72 & 0.33--0.42 & 0.05--0.52\\
LOS, $N_t=2$ & 0.69--0.84 & 2.24--5.69 & censored & 0.02--0.20 & 0.00\\
LOS, $N_t=4$ & 0.45--0.72 & 2.76--5.04 & censored & 0.02--0.17 & 0.02--0.19\\
\bottomrule
\multicolumn{6}{@{}p{0.97\columnwidth}@{}}{\scriptsize Censored: precoded and propagation-channel $B_{c,50}$ both exceed the 78~MHz span.}
\end{tabular}
\end{table}

The per-subcarrier Gram matrix survives precoding. Since
\begin{equation}
\hH\hH^H\propto\mathbf{G}\bH\,\bQ\bQ^H\bH^H\mathbf{G}^H=\mathbf{G}\bH\bH^H\mathbf{G}^H
\label{eq:gram}
\end{equation}
whenever $\bQ\bQ^H=\mathbf{I}$ ($N_t=N_s$) or $\bQ$ spans the row space of $\bH$ ($N_t>N_s=\mathrm{rank}\,\bH$), every function of $\hH(k)\hH^H(k)$ is invariant. This includes the Frobenius norm, the singular values (hence capacity and condition number), and the receive correlation. For $N_t=N_s$, \eqref{eq:gram} is exact even with quantized and stale feedback, because the Givens reconstruction is unitary by construction. In simulation, for $N_t=4$ and in the worst of eight delay-spread and fading cases (300 PPDUs each), the per-subcarrier error of $\|\hH\|_F^2$ has a median (95th percentile) of 0.06 (0.15)~dB with $(6,4)$ feedback, $N_g=4$, and $\rho=0.99$, and 0.51 (1.67)~dB with $(4,2)$, $N_g=16$, and $\rho=0.95$; the capacity at 20~dB SNR changes by 0.5 (0.8)\% and 5.0 (6.2)\%, respectively. Frequency-domain power fading, its spectral spread, and temporal power correlation computed from $\|\hH\|_F^2$ therefore transfer from precoded CSI. Complex-domain per-entry statistics do not: $\tau_{\mathrm{rms}}$, $B_{c,50}$, envelope-based Rician $K$, and phase-based delay or angle estimates per entry.

\section{Measurements}
A commercial $2\times2$ IEEE 802.11be client chipset reported, for each PPDU passing a transmitter-address filter, the CSI of all streams and chains, per-chain RSSI, PPDU format, bandwidth, stream count, and a beamforming flag. Mapped to subcarrier indices, the CSI lay on a 156.25~kHz grid (501 subcarriers at 80~MHz, 1013 at 160~MHz) or a 312.5~kHz grid (114 at 40~MHz, 242 at 80~MHz). AP-A is an IEEE 802.11ax AP (80~MHz at 5~GHz); AP-B is an 802.11be AP (80~MHz at 5~GHz, 160~MHz at 6~GHz); AP-C was run in 802.11n mode (40~MHz) and in 802.11ac mode (80~MHz) at 5~GHz; AP-D is a $4\times4$ 802.11ax AP (80~MHz at 5~GHz). The client was on a desk 2--3~m from AP-A in a laboratory and held in the hand 5--10~m from AP-B in another laboratory. Downlink data came from Internet Control Message Protocol (ICMP) echo at one per second (AP-B, 100~s captures; AP-A, with and without it) or ten per second (AP-C; 18 data and 18 beacon captures of 30~s per mode, the client carried by one of two persons standing or walking at two speeds about 5~m from the AP at three sites, one outdoors). AP-D was captured with beamforming off and on, about 30~s each, the client on a table, handheld, or carried at two walking speeds (beacon RSSI $-38$ to $-44$~dBm), and in a second session on the table at 100 frames/s (three 10-s captures each way). Only two-stream PPDUs were used (Table~\ref{tab:meas}).

\begin{table}[!t]
\centering
\caption{Measured Two-Stream Data PPDUs and Test Outcome ($c_{\mathrm{th}}=0.45$, $q_{\mathrm{th}}=0.9$)}
\label{tab:meas}
\scriptsize
\setlength{\tabcolsep}{2.6pt}
\begin{tabular}{@{}llrcccr@{}}
\toprule
AP & PPDU, band, destination & $n$ & Flag & $\tilde c$, median (IQR) & $q$ & Prec.\\
\midrule
A & HE80, 5 GHz, client & 289 & 1 & 0.24 (0.23--0.24) & 1.00 & 279\\
A & HE80, 5 GHz, client & 2 & 0 & 0.58, 0.59 & 0.76 & 0\\
B & EHT80, 5 GHz, client & 322 & 0 & 0.11 (0.10--0.13) & 1.00 & 318\\
B & EHT160, 6 GHz, client & 283 & 0 & 0.08 (0.08--0.09) & 1.00 & 283\\
B & HE80, 5 GHz, other station & 6 & 1 & 0.75 (0.68--0.79) & 0.38 & 0\\
C & HT40, 5 GHz, client & 5221 & 0 & 0.70 (0.63--0.76) & 0.49 & 0\\
C & VHT80, 5 GHz, client & 5225 & 1 & 0.14 (0.10--0.30) & 1.00 & 4003\\
C & VHT80, 5 GHz, client & 54 & 0 & 0.72 (0.67--0.75) & 0.50 & 0\\
D & HE80, 5 GHz, BF off & 211 & 0 & 0.67 (0.54--0.74) & 0.32 & 0\\
D & HE80, 5 GHz, BF on & 284 & 1 & 0.17 (0.12--0.20) & 1.00 & 271\\
D & HE80, BF off, 2nd & 3012 & 0 & 0.64 (0.64--0.64) & 0.43 & 0\\
D & HE80, BF on, 2nd & 3012 & 1 & 0.21 (0.20--0.22) & 1.00 & 3012\\
\bottomrule
\multicolumn{7}{@{}p{0.97\columnwidth}@{}}{\scriptsize VHT, HE, EHT: very high throughput, high efficiency, and extremely high throughput PPDUs (802.11ac, ax, be); suffix: bandwidth in MHz. Flag: beamforming indication. IQR: interquartile range. $q$: median. Prec.: declared precoded. BF: AP-D beamforming. 2nd: second session.}
\end{tabular}
\end{table}

AP-A set the indication on 289 of its 291 frames; the test declared 279 of these precoded and neither unflagged frame. AP-B set the indication on none of its 605 EHT frames to the client, yet 318 of 322 at 80~MHz and all 283 at 160~MHz tested as precoded. None of the 8452 frames expected to be unprecoded (AP-A's two unflagged frames, as AP-A flagged all others; AP-B's six to another station, precoded for that receiver's channel; AP-C's in 802.11n mode, SNR 22~dB; AP-D's with beamforming off) did at any threshold pair with $c_{\mathrm{th}}=0.35$--0.55 and $q_{\mathrm{th}}=0.8$--0.95, over which detections stay at 880 of the 896 AP-A and AP-B frames to the client and the simulated worst-case false-alarm rate ranges from $2\times10^{-4}$ to $2.2\times10^{-2}$. The 3223 AP-D frames bound the measured false-alarm rate below $9.3\times10^{-4}$ (95\%), and with AP-C's 802.11n frames, 8444 frames bound it below $3.6\times10^{-4}$ if AP-C does not precode in that mode. No unflagged frame of AP-A, AP-C, or AP-D tested as precoded; whether AP-B omits the indication on EHT frames or the receiver drops it, the flag cannot replace the test. The per-PPDU SNR of the CSI (delay-domain noise floor, all entries, as in simulation) had medians of 27--29~dB (5th percentile ${\ge}24$~dB), where fresh feedback gives $P_D\ge0.98$. In 802.11ac mode, AP-C flagged 5225 of its 5279 frames; the test declared 4003 of these precoded (79--100\% per capture standing, 65--74\% walking) and none of the other 54. Of the 16 AP-A and AP-B frames not declared precoded, two are the unflagged AP-A frames; the other 14 failed the coherence condition (median $\tilde c$ of 0.73 and 0.75), as did 12 of AP-D's 13 misses and 1186 of AP-C's 1222, mostly while walking (10 and 1097); at median SNRs of 23--37~dB, this indicates imperfect steering rather than noise (Table~\ref{tab:det}).

At $\mathbf{D}=\mathbf{I}$, RSSI-based per-chain equalization raised the median coherence of the precoded AP-A and AP-B groups from 0.14--0.29 to 0.38--0.77; with the minimization, no decision changed. The minimizing ratio had medians of $-1$ to $+1$~dB for AP-A to AP-C, as for a beamformee that computes $\bV$ from its unequalized estimate, but $-6$~dB at AP-D's table placement in both sessions (cause unknown; sounding frames were not captured), where the test at $\mathbf{D}=\mathbf{I}$ detected 32 of 72 and 0 of 3012 frames (minimized: 71 and 3012). In every group, 90\% of them lay within $-8$ to $+6$~dB (one of 17{,}921 frames at the $\pm15$-dB limit; $\pm20$~dB leaves simulated false-alarm rates unchanged).

In the measured data, averaged over chains (re-ordered where the client had exchanged them; the symmetric gain search leaves the test unchanged), the stream-2 to stream-1 $\tau_{\mathrm{rms}}$ ratios are 1.96, 1.88, and 2.05 in the three precoded groups of AP-A and AP-B. Unprecoded columns gave ratios of 0.87--1.19 (AP-C, 802.11n) and 1.04--2.60 (AP-D, beamforming off); precoded, AP-C gave 1.40--1.86 over its primary 40~MHz (the 802.11n channel), but AP-D 0.94--1.35, as Section~IV allows for one placement, so the ratio separates neither way. On the same 40~MHz, spot, and motion, AP-C's per-stream $\tau_{\mathrm{rms}}$ in 802.11ac mode was 0.22--0.83 times that in 802.11n mode, while its beacons changed by a median of 9\% (at most 25\%). With beamforming off, 80--100\% of AP-D's stream-2 impulse responses (stream~1: 0--50\%; beamforming on: at most 7\%) show a copy 38--63~ns from the main path within 10~dB, as from two antennas per stream with cyclic shifts 50~ns apart. Switching beamforming on removed that copy and lowered both streams' $\tau_{\mathrm{rms}}$ by 31--56\% with the client handheld, while the beacons changed by $-8$ to $+9\%$. They spread over 53--87~ns, against 18--37~ns for unprecoded data streams on the same 20~MHz, close to the 59--67~ns that \eqref{eq:cdd} gives for AP-D's four chains at equal power.
\section{Conclusion}
From the CSI alone, the gain-adjusted test found precoding in 77--100\% of the two-stream frames each AP beamformed to the client and in none of the unprecoded ones. For CSI-based sounding: run the test on every multi-stream capture and treat a frame as precoded if the test or the beamforming flag says so; for precoded captures, report only statistics of $\hH\hH^H$, or obtain unprecoded CSI from null data packets, non-HT frames, or a transmitter with beamforming off; treat a negative test at low SNR, on a nearly rank-one link, or with the client in motion as inconclusive; and account for transmit cyclic shifts in non-HT and spatially expanded CSI before comparing it with models.

\section*{Acknowledgment}
The authors used Claude Opus 5.5 (Anthropic, Nov. 2026) to help draft and edit the text and to write the analysis and figure code; the authors checked all results, equations, and references.

\bibliographystyle{IEEEtran}
\bibliography{refs}

\end{document}